\def\deg{{^\circ}}

\def\la{\mathrel{\hbox{\rlap{\hbox{\lower4pt\hbox{$\sim$}}}\hbox{$<$}}}}
\def\ga{\mathrel{\hbox{\rlap{\hbox{\lower4pt\hbox{$\sim$}}}\hbox{$>$}}}}

\def\arcmin{\hbox{$^\prime$}}

\newcommand{\msun}{\,{\rm M}_\odot}

\documentclass{iaus}
\usepackage{graphicx}

\title 
[Galaxies in the Zone of Avoidance]%% give here short title %%
{Highlights of multi-wavelengths surveys\\in the Zone of Avoidance}

\author[Kraan-Korteweg et al.]%% give here short author list %%
{Ren\'ee C. Kraan-Korteweg$^1$,\\
      Kurt J. van der Heyden$^1$,
      Michelle E. Cluver$^1$,
 \and Patrick A. Woudt$^1$}

\affiliation{$^1$ Astronomy Department, University of Cape Town\\
             Private Bag X3, Rondebosch 7701, South Africa\\
email: {\tt kraan@ast.uct.ac.za,} \\
          {\tt michelle@ast.uct.ac.za,}   
          {\tt heyden@ast.uct.ac.za,}   
          {\tt pwoudt@ast.uct.ac.za}}
\pubyear{2009}
\volume{xxx}  %% insert here IAU Symposium No.
\pagerange{1--8}
\jname{First Middle East-Africa Regional IAU Meeting}
\editors{Ahmed Hady ed.}

\begin{document}

\maketitle

\begin{abstract}
Rather than giving a complete overview on extragalactic Zone of
Avoidance (ZOA) research, this paper will highlight some interesting
discoveries in the ZOA, such as new near- to far-infrared observations
(IRSF, {\sl Spitzer}) of the most massive disk galaxy found to-date
(HIZOA 0836-43), and deep multi-wavelength observations of a spiral
galaxy WKK~6167 undergoing transformation while infalling along the
Great Attractor Wall into the Norma cluster -- reminiscent of similar
incidences observed in only two galaxies at higher redshifts ($z \sim
0.2$; Cortese et al. 2007). While the recent systematic
multi-wavelengths approaches to uncover the large-scale structure of
galaxies across the ZOA have proven quite successful, in particular in
the Great Attractor region, they lack the required depth to answer
open questions with regard to our understanding of the dynamics in the
Local Universe. The actual mass distribution is poorly understood and
does not satisfactorily explain the observed peculiar velocity fields
and the CMB dipole. We will present future HI survey strategies to be
pursued with the South African SKA Pathfinder MeerKAT in the ZOA, that
can -- amongst others -- resolve the long-standing Great
Attractor/Shapley controversy, and determine at which
distance range the cumulative peculiar motion of the Local Group
flattens off and the Universe becomes homogeneous.

\end{abstract}


The dust, stars and gas in the plane of the Milky Way result in a
Zone of Avoidance (ZOA) in the extragalactic sky, the size and shape of
which depends on the wavelength. An unbiased "whole-sky" map of galaxies is
essential, however, for understanding the dynamics in our local
Universe, i.e. the peculiar velocity of the Local Group (LG) with
respect to the Cosmic Microwave Background (CMB) and velocity flow fields
such as in the Great Attractor (GA) region. 

Enormous progress has been achieved in the last 20 years in narrowing
this ZOA through various systematic observational multi-wavelength
surveys --- and sometimes serendipitous discoveries. These efforts and
subsequent results were reviewed in great detail in Kraan-Korteweg \& Lahav
(2000) with an update by Kraan-Korteweg in 2005, and will not be
repeated. This paper will focus on some particularly interesting
objects discovered behind the Milky Way next to future 
perspectives for ZOA research.

\section{A supermassive HI galaxy}

The galaxy HIZOA~J0836-43 was first discovered in the deep Parkes (64m
single dish telescope) MultiBeam (MB) systematic HI-survey of the
southern Galactic Plane ($|b| \le 5\deg$ (see Kraan-Korteweg 2005,
Kraan-Korteweg et al. 2005, for preliminary results). It was also
identified in the more shallow HIPASS survey, the systematic all sky
HI Parkes All Sky (southern) Survey performed with the same instrument
(Meyer et al. 2004). It is one of the most HI-rich galaxies ($M_{\rm
HI} = 7.5 \cdot 10^{10}\msun$). Unlike other giant HI galaxies, (like
e.g. Malin~1; Bothun et al. 1987; Pickering et al. 1997) this disk
galaxy is not of low surface brightness. According to hierarchical
structure formation they are only forming now ($z < 1$; Mo et
al. 1998). Such massive spiral galaxies are extremely scarce (~$1/3
\cdot 10^7$ Mpc$^3$). At $v=10700$km/s ($D=148$Mpc), HIZOA J0836-43 is
the nearest of its kind. However, a detailed analysis is hampered by
its location behind the Milky Way: we view it through ~10mag of
extinction ($A_B$) and only observations in the near-, mid- and
far-infrared will prevail.

New results about his enigmatic galaxy were derived based on deep near
infrared ($JHK$) images (IRSF, SAAO), mid- to far-infrared images and
high resolution spectroscopy (IRAC, MIPS and IRS; {\sl Spitzer} Space
Telescope). A more detailed analysis of this galaxy based on this new
data is given in Cluver et al. 2008. Further results will be detailed
in forthcoming papers (Cluver et al. 2009, in prep., UCT PhD thesis,
2009).

\begin{figure}
\vspace*{-0.7 cm}
\begin{center}
 \includegraphics[width=3.4in]{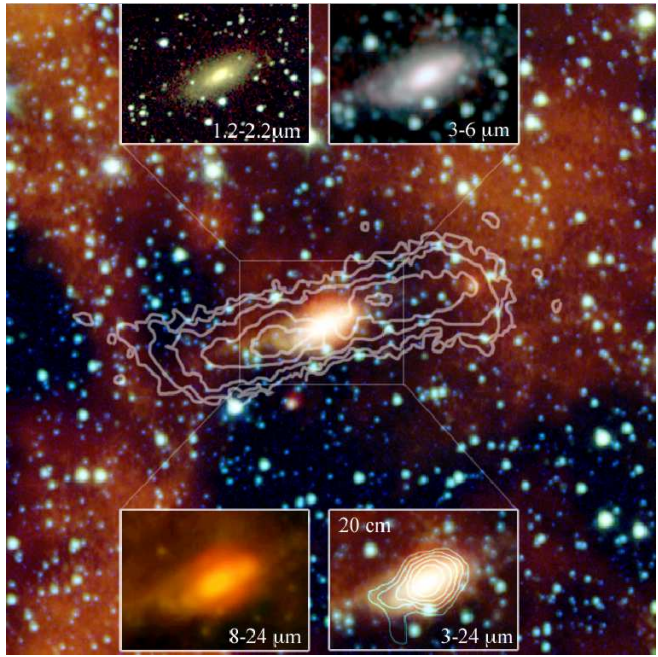}
% \vspace*{-1.0 cm}
 \caption{Infrared view of HIZOA J0836-43 through the
Vela region of the Milky Way. The main image (FOV $\sim$4';
$\sim$170kpc) is a composite of the NIR $J,H,K$, the Spitzer IRAC (3.6,
4.5, 5.8, 8.0$\mu$m) and MIPS (24$\mu$m) bands. The contours reflect
the extended HI distribution (levels  0.1, 0.2, 0.4, 0.8, 1.2, and 
1.6 Jy\,beam$^{-1}$km/s; Donley et al. 2006). The insets
($\sim$1') display various composite images; the bottom right 
3 - 24$\mu$m image, related to starformation, also shows 
the 20cm radio contours.  Figure from Cluver et al. (2008).}
\label{MIR_massive}
\end{center}
\end{figure}

Figure~\ref{MIR_massive} presents an infrared view of HIZOA
J0836-43. The HI contours (Donley et al. 2006) demonstrate the
enormous size of the HI disk, while the galaxy's bolometric luminosity
largely arises from infrared radiation, particularly at longer
wavelengths.  The $1-5\, \mu$m window traces the old stellar
population (top panel, Fig. \ref{MIR_massive}) and the galaxy appears
as an inclined extended disk galaxy with a prominent bulge population.
Its light distribution indicates an early Hubble Type of
$\sim$S0/Sa. The MIR ($\lambda > 5\, \mu$m) is sensitive to the
interstellar medium, notably the warm dust continuum and emission from
Polycyclic Aromatic Hydrocarbon (PAH) molecules. PAHs produce broad
emission bands in the MIR linked to ongoing or recent star formation
(Allamandola et al. 1985).

The $8-24\, \mu$m composite shows strong emission from PAH molecules
and warm dust. The 20 cm radio continuum is correlated with the
$8-24\,\mu$m emission, indicating a common star-formation
origin. While the bulk of the infrared emission is concentrated in the
central nuclear region, it is clearly extended and exhibits spiral-arm
asymmetries.

In summary, the near-infrared properties of the galaxy appear typical
for an S0 system, while radio observations suggest recent active star
formation. The {\sl Spitzer} observations find HIZOA J0836-43 to be a
luminous infrared starburst galaxy with a star formation rate of
$~21\msun/$yr, arising from exceptionally strong molecular PAH
emission and far-infrared emission from cold dust. However,
high-resolution spectroscopy of the galaxy reveals a weak mid-infrared
continuum compared to other starforming galaxies and U/LIRGs.

An accompanying deep near-infrared survey (2.2 $\Box^\circ$) finds
this HI-rich galaxy to lie in a slightly enhanced density of fainter
galaxies within a (newly identified) filamentary structure encircling
a void (Cluver 2009). The conditions appear favourable for the galaxy
to have grown this large HI disk, both through accretion of gas due to
minor merging, as well as infall of gas along the filament. It also
seems consistent with the observation that HI-massive galaxies are
preferentially found in low density regions.

\section{Mapping the Great Attractor}

Following our deep optical galaxy search in the Great Attractor (GA)
region, and the recently completed HIZOA 21-cm radio survey which
encompassed the GA region, a clear outline was unveiled of a
large-scale structure running nearly parallel to the Galactic plane at
the distance of the GA. This large supercluster of galaxies, dubbed
the Norma wall, has been the subject of a deep near-infrared follow up
survey (44$\Box^\circ$) with the IRSF ($JHK_s$), in order to gauge its
dynamical contribution to the overall galaxy flow in the GA region. At
the core of the Norma wall, the rich and massive Norma cluster resides
(Kraan-Korteweg et al. 1996, Woudt 1998). A dynamical analysis of
this cluster has revealed various (infalling) subgroups (optical and
X-ray) and a number of dynamically-peculiar galaxies in this clusters
have been identified (Woudt et al. 2008). A particularly interesting 
one is disucssed below.

\subsection{A galaxy in transformation in the Norma cluster}

Recently, the presence of a 70-kpc long X-ray tail and a 40-kpc
H$\alpha$ tail, respectively, was reported by \cite{sun06,sun07}, in
WKK\,6176, a spiral galaxy in the nearby ($cz = 4871 \pm 54$ km/s) and
massive ($M_{R < 2\,{\rm Mpc}} = 1 \times 10^{15}$ M$_{\odot}$) Norma
cluster, \cite{woudt08}. This galaxy is the low-redshift equivalent of
the two recently detected spiral galaxies in massive rich clusters
(Abell 2667 and Abell 1689) at $z \sim 0.2$ which show clear evidence
for strong galaxy transformation, \cite{cort07}.

\begin{figure}
\begin{center}
 \includegraphics[width=3.4in]{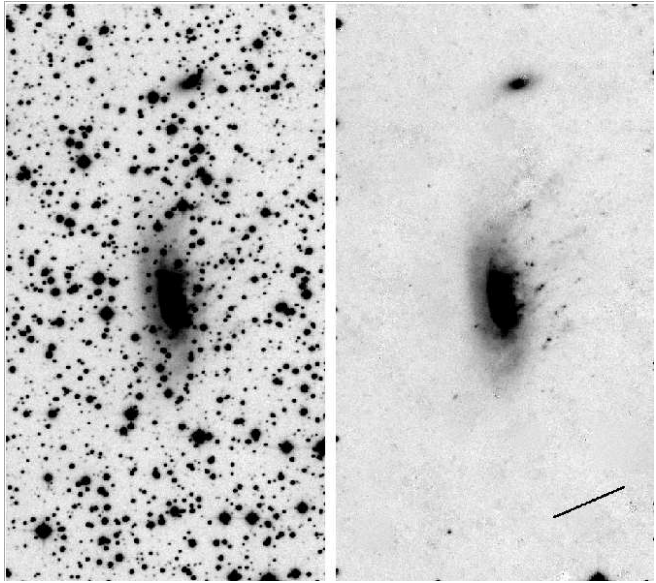}
\caption{A deep $R_C$ image of WKK\,6176 ($2.2' \times 4.0' = 43
\times 77$ kpc).  The left image shows the original data, the right
image shows WKK\,6176 after star subtraction using the KILLALL routine
\cite{buta99}. Low surface brightness filaments and bright knots are
clearly visible.}
\label{jelly}
\end{center}
\end{figure}

The X-ray tail of WKK\,6176 is aligned with the major axis of the
galaxy-density profile of the cluster which is indicated by the
diagonal line in the right panel of Fig.~\ref{jelly}, which itself
is aligned with the main large-scale structure of the Norma Wall,
\cite{woudt08}. Fig.~\ref{jelly} shows our deep $R_C$ image of
WKK\,6176 before and after star-subtraction; it demonstrates the
effectiveness of the star-subtraction. Numerous low-luminosity
filaments and bright knots (not foreground stars) stand out, giving
the galaxy its `jelly-fish' appearance.

Given the proximity of the Norma cluster, WKK\,6176 provides an
excellent opportunity to study the interaction of a galaxy with the
intracluster medium (ICM) at high resolution and sensitivity. Deep $B
V R_C J H K_s$ photometry of WKK\,6176 (already obtained) will be used
to generate pixel-by-pixel colour-magnitude diagrams and colour-colour
diagrams \cite{grijs03} to study the star formation history of
WKK\,6176 in combination with GALEV, \cite{bicker04}.

The GALEV models include an ever growing grid of refined models of
undisturbed Sa, Sb, Sc and Sd galaxies falling into a cluster
environment at a wide range of redshifts and experiencing various
star formation scenarios (e.g., starbursts of varying
strengths and time scales). For all those models, the
evolution of the galaxies' spectra (ultraviolet, optical and
near-infrared) and broad band spectral energy distribution is
determined.

WKK\,6176 is the nearest galaxy observed in a state of strong
transformation through visible interactions with the ICM of a rich and
massive cluster. Based on our multiwavelength observations and a
comparison with galaxy evolution synthesis models, we aim to constrain
the recent star formation history.

\section{Future HI-surveys and the Great Attractor/Shapley Controversy}

HI-surveys have proven the most effective in tracing (HI-rich)
galaxies in areas of extreme star crowding and dust column
density. They suffer no significant selection effects. However, even
the deepest systematic HI ZOA surveys to date (Parkes MB;
e.g. Kraan-Korteweg et al. 2005) remain very shallow, with e.g. a sky
density of only less than 1 gal$/\Box^\circ$ in the GA overdensity. It
does not probe deep into the GA Wall (mostly $\ga M^{*}_{\rm HI}$
galaxies). Deeper surveys would cause confusion problems due the
limited spatial resolution of the Parkes MB ($4\arcmin \times 4\arcmin
\times 26$km/s). The same holds for HIMF derivations. This low
resolution does not permit a differentiation of low mass
objects/satellites from neighbouring larger spirals, resulting in an
underestimate of the number of low mass galaxies, while overestimating
the mass of large galaxies. Covering a larger (or higher) redshift
range would suffer such confusion problems even stronger. This is
particular relevant in trying to solve the long-standing Great
Attractor/Shapley controversy: who is the major attractor in the local
Universe? Is the Shapley cluster concentration (SH) at 15000km/s the
dominant contributor to the dipole motion of the LG (e.g. Saunders et
al. 2000; Basilakos \& Plionis 2006; Kocevski et al. 2005), or does
the cumulative velocity level off beyond $v \ga 5000$km/s
(e.g. Erdogdu et al. 2006a,b).

Improved sensitivity is required ($\sim 2$ orders of
magnitude), better spatial resolution (at least $0.5-1\arcmin$), wider
instantaneous bandwidth, increased number of channels, and a large
FOV for survey speed. These ambitions will be achieved with the SKA
Pathfinders. In the following, a survey strategy with MeerKAT, the
South African Pathfinder (see http://www.ska.ac.za), will be
presented to map the GA/Shapley overdensities.

\subsection{MeerKAT, the South African SKA Pathfinder} 

MeerKAT will be constructed near Carnavoron in the
Northern Cape (SA). A high speed data transfer network will link the
telescope site in the Karoo to a remote operations facility.  
KAT-7, a 7-dish engineering testbed and science instrument will be
commissioned towards the end of 2009. The full array of 80 or more
dishes should be ready by 2012.  It is anticipated that
science projects will be performed at some intermediate stage to test
and optimise the running of the telescope. Hence our simulations
include calculations not only for a 7 and 80 dish array
configuration, but also for 30.

At the time of the meeting (April 2008) the MeerKAT instrument
specifications were 80 x 12m dishes, with cooled receivers (30K),
single pixel feed, 2 polarisations, $1\deg$ beamsize, and a FOV of
0.8$\Box^\circ$. KAT-7 will have 256KHz instantaneous bandwidths, the
full array 512MHz, possibly 1024 MHz. The frequency range will be
1.2-2.0 GHz for KAT-7 and $0.5-2.5$GHz for MeerKAT (possibly
$0.3-3$GHz).  The baseline for KAT-7 will be short (210m), but will extend
to 10km for MeerKAT, with a dense core (70\% within 700m). It will have
16k channels, allowing for excellent velocity and linewidth
resolution.

\subsection{Survey assumptions}

The minimal survey region in the ZOA that will address both the mass
and mass distribution in the GA Wall, as well as the suspected
SH-extension across the ZOA --- which at low Galactic latitude lies
conveniently (or rather more like a conspiracy) exactly behind the
GA-Wall --- is of the order of 200$\Box^\circ$ ($312\deg \le \ell \le
332\deg; |b| \le 5\deg$). This area overlaps with an ongoing deep
near-infrared ($JHK_s)$ survey using the IRSF, as well as a smaller
areal survey with the {\sl Spitzer} mid-infrared IRAC imager, an
international collaboration including Riad (UCT), Jarrett
(CalTech), Nagayama (Kyoto) and Wakamatsu (Gifu).

We demand S/N$ = 5$ for robust detection, and assume a channel width
of 15km/s. We apply optimal smoothing to the number counts,
i.e. binning over the line width to optimise the signal-to-noise. The
scientific assumptions are based an HI mass function (HIMF) as
determinated in Zwaan et al.~(2005) derived from the systematic HI survey
HIPASS (N$=4315$; Meyer et al. 2004). Their HIMF was fitted with a
Schechter function. The resulting parameters are $\alpha = -1.37$ for
the faint end slope, log($M^{*}_{\rm HI}) = 9.80\msun$ for the
characteristic HI mass, and $\Omega{^{*}} = 6.0 \cdot 10^{-3}$
Mpc$^{-3}$dex$^{-1}$ for the volume density. Their results are in good
agreement with the HIPASS Bright Galaxy Catalogue, a complete but
smaller, sub-sample of galaxies (Zwaan et al. 2003). We adopt the same
cosmological parameters of H$_0=75$Mpc/km/s, $\Omega_m = 0.3$, and
$\Omega_{\Lambda}=0.7$.
 
Interestingly, they find tentative evidence for environmental effects:
the HIMF becomes steeper towards higher density regions and lower in
lowest density environments ($\alpha=-1.5$, $-1.2$
respectively). Considering that we are regarding high density areas,
we adopt this steeper slope ($\alpha = -1.5$) for the redshift range
of the GA and SH regions, i.e. $3500 < v < 6500$km/s and $13500 < v <
18000$km/s. We furthermore adopt an overdensity of $\rho / <\rho>=3$
and 9 for the GA and SH distance intervals. The first value is close
to observational data for that area, while the latter is a minimal
assumption, given that some papers (e.g. Kocevski et al. 2005) seem to
indicate that the SH region, at 3 time the distance of the GA, is
exerting the same gravitational attraction on the LG as the GA
overdensity.

\subsection{The simulations}

\begin{figure}
\begin{center}
\includegraphics[scale=0.55,angle=-90]{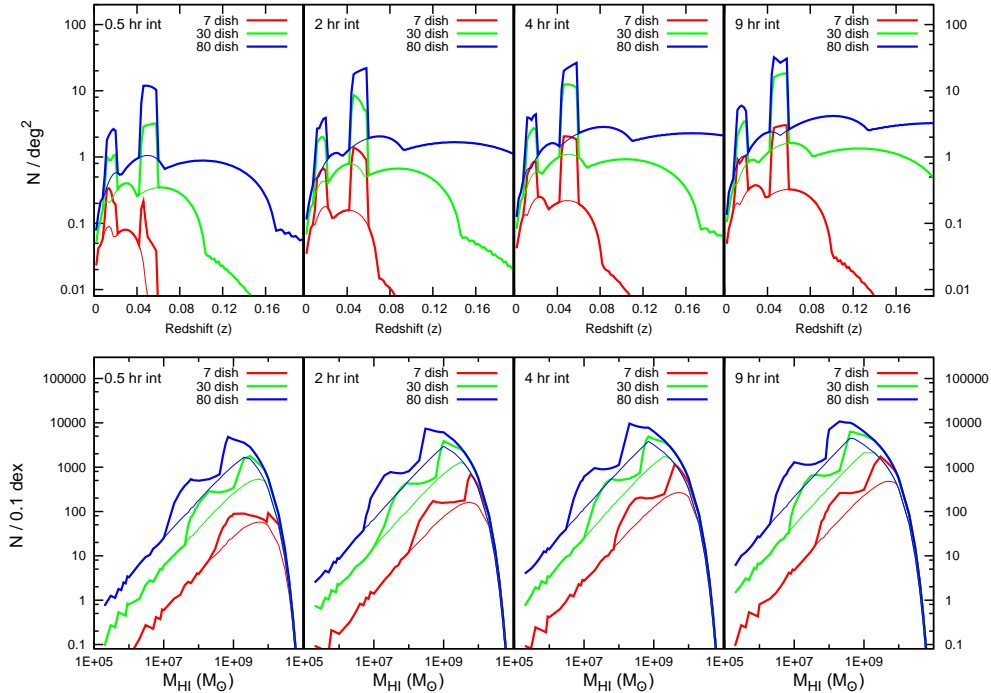}
\caption{Various MeerKAT survey scenarios assuming overdensities of 3
and 9 in the GA and SH region respectively. The figure displays
projected sky density versus redshift (top panel) and log HI-mass
versus redshift (bottom panel) for integration times of 0.5, 2, 4, and
9hrs.}
\label{simu}
\end{center}
\end{figure}

Results based on these simulations are illustrated through a variety
of plots in Fig.~\ref{simu} and some characteristic parameters in
Table~1.  The top panel of Fig.~\ref{simu} displays detections as
number $N/\Box^\circ$ for increasing redshift $z$ in intervals of
$\Delta z = 0.002$ (corresponding to $\Delta v = 600$km/s) for
integration times of 0.5, 2, 4, and 9 hours for the different KAT
phases (7, 30 and 80 dishes). The bottom panel shows the HI mass
distribution, where the numbers $N$ are given in steps of
0.1dex~log(M$_{\rm HI}$). The interpretation of these plots should be
made in conjunction with Table~1, which lists the duration of the
survey given the assumed integration times per pointing for the
several KAT versions, resulting number density $N/\Box^\circ$ as well
as the HI-mass limit (in log) for the GA and SH overdensities.
 
It is reassuring to note that the prediction based on the simulation
for the GA region of the 30-dish configuration with a 0.5hr
integration is in close agreement with observational results from the
Parkes ZOA HI survey of that same area, namely 0.8$/\Box^\circ$ for
the 25 min integrations of HIZOA, and a system temperature ($\sim$30K)
and aperture that closely correspond to the KAT-30 instrument
specifications.

\begin{table}
\begin{center}
\includegraphics[scale=0.6]{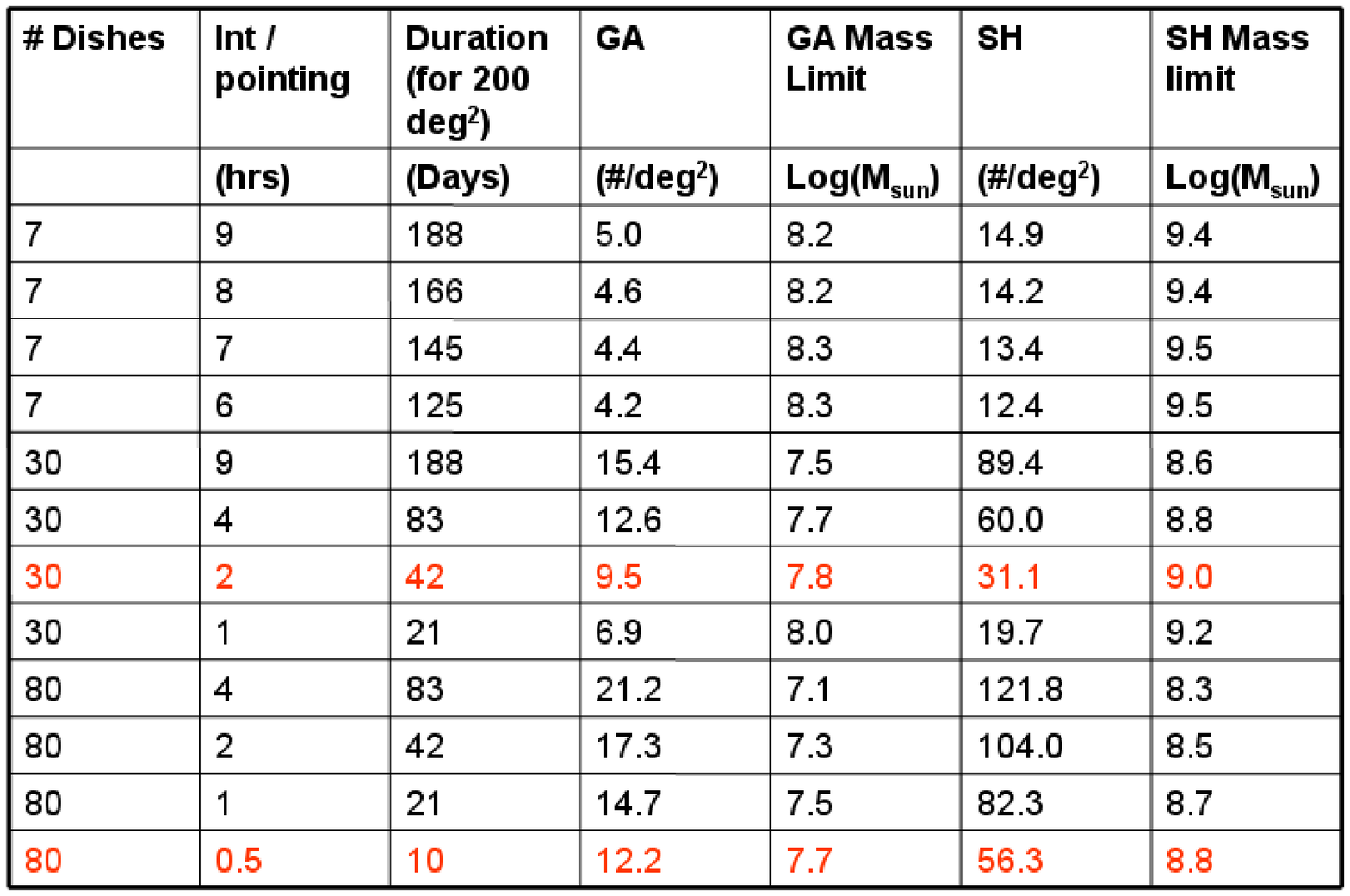}
\caption{Integration times per pointing for various KAT operation phases 
for various observing periods (in days) to attain reasonable number 
density and HI mass limits for both the GA and Shapley distance range.}
\label{scenarios}
\end{center}
\end{table}

\subsection{Conclusions}
It is notable that all simulations predict a signal of these
overdensities to be quite prominent in the distribution. Despite these
clear peaks in the distributions, it is evident that KAT-7 is not a
suitable option to take this project a step further. The resulting
detection rate will remain quite low even for the longest regarded
integration time of 9 hrs (requiring half a year of observing time):
only a few dozen high-mass galaxies in the SH region - although their
detection would immediately indicate an overdensity (compare thick to
thin line in Fig.~\ref{simu}), and at most a few more low mass galaxies
in the GA region.

However, the 30-dish configuration will allow already quite
significant results with, e.g., the 2hr integration time (requiring a
total survey period of 42 days). Such a survey will trace galaxies
into the dwarf regime (7.8\,log$\msun$) in the GA region with a good
number density (though not too dense to run into confusion problems),
and will satisfactorily trace galaxies in the SH region close to one
order of magnitude below the characteristic HI-mass of
9.8\,log$\msun$. The number densities are $\sim 31/\Box^\circ$ --
``if'' the SH overdensity extends across the Galactic Plane and
connects to the massive X-ray clusters Triangulum Australis cluster
and CIZA~J1652 at $b \sim -10\deg$. These clusters, together with
CIZA~J1514 and CIZA~J1518 above the plane ($b\sim+10\deg$), give rise
to a large fraction of the steep dipole increase at around the SH
cluster concentration distance located at slightly higher latitudes
(see Kocevski et al. 2005; Fig.~3). The full MeerKAT operations allow
such a survey in only 10 days of half hour integration.

We can maintain that KAT-30 would be the ideal interim phase to
conclusively resolve the GA/SH controversy for the first time while
the full MeerKAT configuration the ideal survey instrument to
extend this pilot GA/SH survey to the whole southern ZOA. The latter
would require 90 days. An extension to higher latitudes around the
Galactic Bulge would also be recommended, as mapping of extragalactic
large-scale structures is limited there, even at medium extinction
levels (Kraan-Korteweg et al. 2008).

Apart from being an ideal probe to track the large-scale structures of
the mass distribution in the GA and SH regions, the proposed 2hr
integration, 30-dish MeerKAT survey would provide a deeper insight into
the HIMF, particularly with regard to the low mass end, which is still
ill-constrained.  Moreover, it will allow further insight into
environmental effects, as we will probing high density areas that form
part of the both GA/SH overdensity, wall-like features that seem to
connect these mass densities across the ZOA, as well as the substantially
underdense spaces (voids) inbetween the GA overdensity and the
suspected Shapley Wall (see e.g. Woudt et al. 2008).

\medskip

\noindent{\bf Acknowledgements.} -- RKK greatly appreciates the grant
allocated by the IAU to attend this meeting. Financial support from
the National Research Foundation (RKK, MCC, PAW) and South African SKA
Human Capacity Development Programme (KvdH) is gratefully
acknowledged.

\end{document}